\documentclass{sf2a-conf2024}
\usepackage{graphicx}
\usepackage{hyperref}
\usepackage[]{natbib}  
\usepackage{epstopdf}

\def\BibTeX{{\rm B\kern-.05em{\sc i\kern-.025em b}\kern-.08em
    T\kern-.1667em\lower.7ex\hbox{E}\kern-.125emX}}
\bibpunct{(}{)}{;}{a}{}{,}  

\begin{document}

\TitreGlobal{SF2A 2024}


\title{Solar p-mode excitation rate along the magnetic activity cycle}

\runningtitle{Short title here}

\author{E. Panetier}\address{Université Paris Cité, Université Paris-Saclay, CEA, CNRS, AIM, 91191, Gif-sur-Yvette, France}

\author{S. N. Breton}\address{INAF – Osservatorio Astrofisico di Catania, Via S. Sofia, 78, 95123 Catania, Italy}

\author{R. A. García}\address{Université Paris-Saclay, Université Paris Cité, CEA, CNRS, AIM, 91191, Gif-sur-Yvette, France}

\author{A. Jiménez$^{4,}$}\address{Instituto de Astrofísica de Canarias (IAC), 38205 La Laguna, Tenerife, Spain Universidad de La Laguna (ULL), Departamento de Astrofísica, 38206 La Laguna, Tenerife, Spain}\address{Universidad de La Laguna (ULL), Departamento de Astrofísica, 38206 La Laguna, Tenerife, Spain}

\author{T. Foglizzo$^2$}

\setcounter{page}{237}


\maketitle


\begin{abstract}
Magnetic cycles of solar-like stars influence their internal physics. Thus, the frequency, amplitude, excitation rate, and damping of the acoustic oscillation modes ($p$\,modes) vary with the cycle over time. We need to understand the impact of magnetic activity on $p$\,modes in order to characterise precisely stars that will be observed by the ESA PLATO mission, to be launched late 2026 with the objective to find Earth-like planets around solar-type stars. In this work, we investigate the variation of mode excitation in the Sun during Cycles 23, 24 and the beginning of Cycle 25. To do so, we analyse data obtained since 1996 by two instruments onboard the SoHO satellite: the GOLF spectrometer and the VIRGO sunphotometer. We use a method enabling us to reach a better temporal resolution than classical methods. Combining the variations of energy for several modes $l=[0-2]$ in three frequency bands (i.e. [1800, 2450], [2450, 3110], [3110, 3790]\,$\mu$Hz), our preliminary results show that more energy is associated to several modes during cycle minima, suggesting that there could be a second source of excitation other than turbulent convection that would excite several modes at a time during solar minima.
\end{abstract}

\begin{keywords}
asteroseismology, helioseismology, stellar activity, solar-type, oscillations, data analysis
\end{keywords}


\section{Introduction}
The Sun's convective envelope generates, by dynamo effect, a mean surface magnetic field whose strength evolves on an 11-year cycle, with a change in polarity at the end of each cycle. Similar activity cycles exist in other solar-type stars, influencing their dynamics, including the properties of acoustic oscillations ($p$\,modes), excited by convection. In particular, magnetic cycles have an impact on $p$\,modes properties such as frequency, amplitude, excitation and damping \citep[e.g.][]{broomhall_suns_2014}. In the case of the Sun, $p$\,mode frequencies increase with the cycle strength whereas their amplitudes decrease. Although the frequency shift behaviour has already been extensively studied, we still need to investigate the impact of the magnetic cycle on excitation properties of the modes \citep{jimenez-reyes_excitation_2003,kiefer_they_2021}. Our objective is to investigate the variation of mode excitation following in \citet{foglizzo_are_1998-1} which allows a better resolution in time and in frequency than classical methods. It consists in the reconstruction of a time series for each mode. In the Fourier domain, a $\Delta\omega$ window centred on the mode is selected and the inverse Fourier Transform is computed, allowing to reach a temporal resolution $\delta t = \frac{1}{\Delta \omega}$.


\section{Helioseismic time series analysis}

\begin{figure}[!ht]
    \centering
    \includegraphics[width=0.8\linewidth]{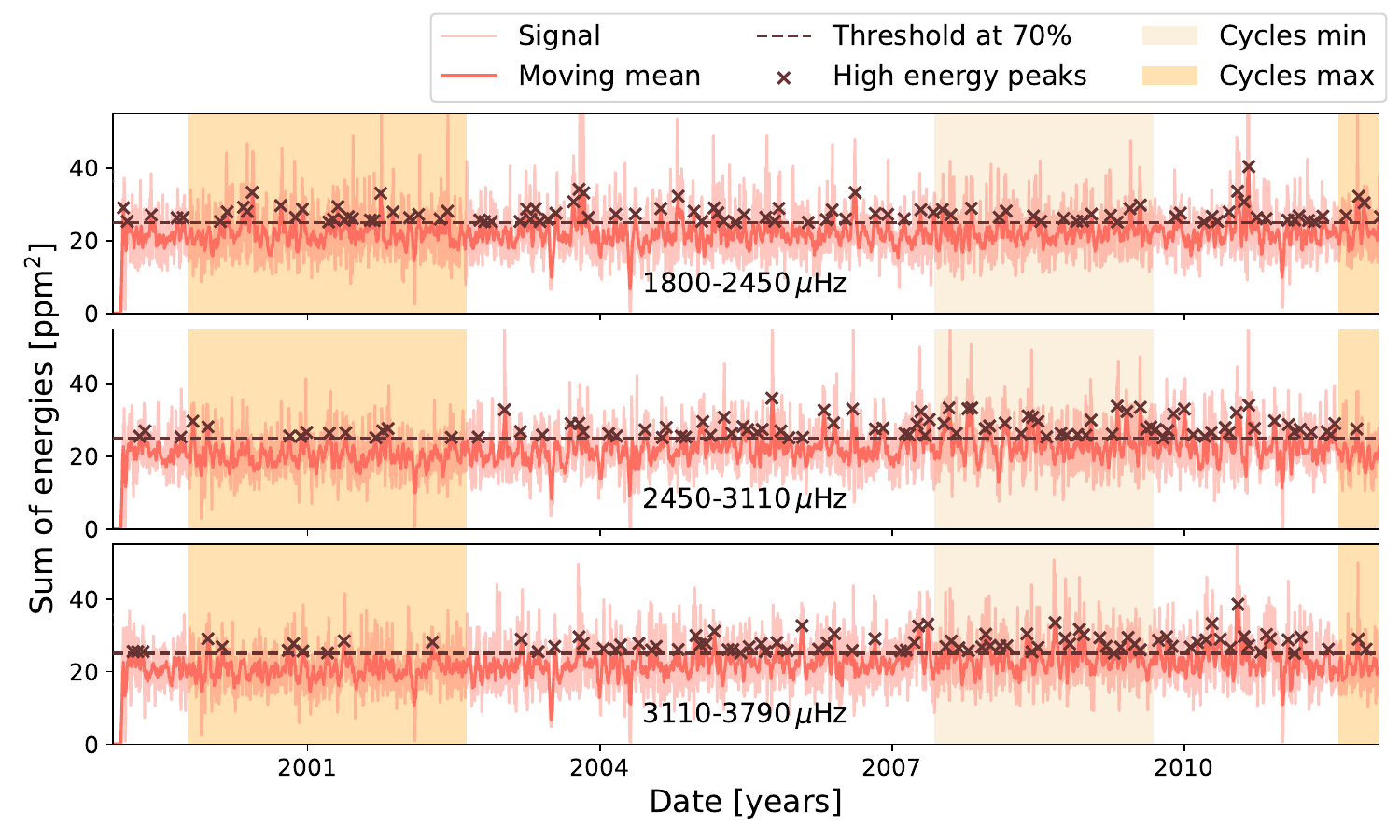}\hfill
    \includegraphics[width=0.8\linewidth]{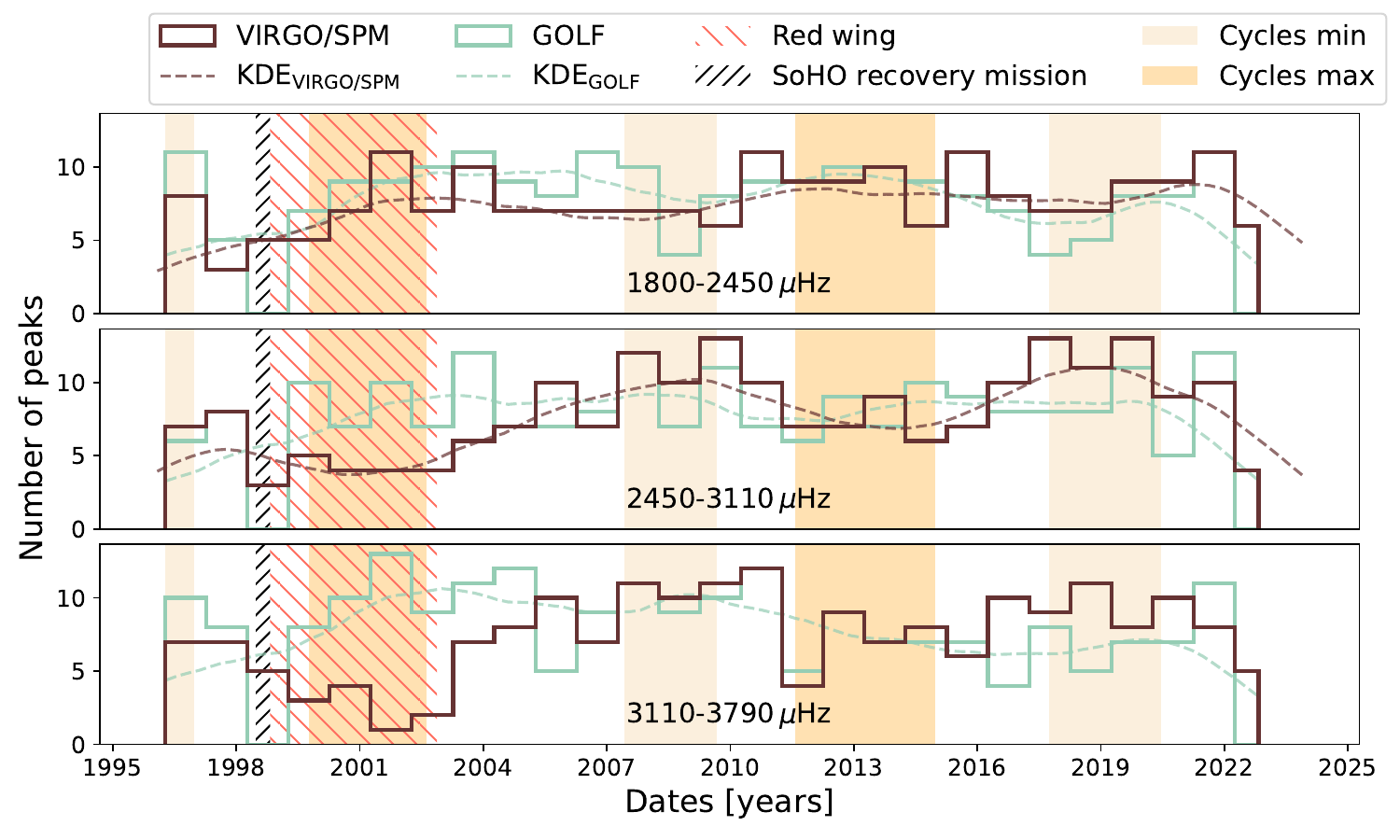}
    \caption{\textbf{Top:} Combined modes reconstructed time series of VIRGO/SPM in three frequency ranges (light red), and their mean over a 7-points moving window (red). The dashed brown line shows the 70\% quantile of the time series and the counted high peaks above it are marked by brown crosses. Yellow zones show dates of maximum (dark) strength of Cycle 23 and beginning of Cycle 24, and minimum (light) strength of Cycle 24. \textbf{Bottom:} Repartition of the higher energy peaks in the combined normalised time series from GOLF (green) and VIRGO/SPM (brown) over time. Top to bottom panels stands for combined signals in the three frequency ranges. Dashed curves show Epanechnikov Kernel Density Estimation (KDE) of each histogram. Yellow zones show dates of minimum (light) and maximum (dark) cycle strength.}
    \label{fig:figure}
\end{figure}

We used data of two instruments of the Solar and Heliospheric Observatory \citep[SoHO,][]{domingo_soho_1995} mission launched in December 1995, i.e. the Global Oscillation at Low Frequencies \citep[GOLF,][]{gabriel_global_1995, garcia_global_2005} and the Variability of solar IRradiance and Gravity Oscillations-Sun PhotoMeters \citep[VIRGO/SPM,][]{frohlich_virgo_1995}. Time series of all modes from $n=14$ to 25 and $l=0$ to 2 were reconstructed with the previously mentioned method by \citet{foglizzo_are_1998-1}. The chosen window size is 8\,$\mu$Hz around each mode whose frequency $\nu$ in the Fourier domain is taken from peakbagging made by \citet{lazrek_first_1997}. The energy of each mode is computed as $E_{n,l} = 2|f_v(t)|^2$, where $f_\nu(t)$ is the inverse Fourier Transform of the filtered mode. The resulting resolution of the time series is $\delta t \sim 1.45$\,days and we later work with mean-normalised energy times series. According to Kolmogorov-Smirnov comparisons, each reconstructed time series agrees well with an exponential distribution as expected \citep{foglizzo_are_1998} for fully randomly excited modes. 

$P$\,modes are more sensitive to the most external regions of the star: magnetic field variations in these regions should influence more modes behaviour \citep{christensen-dalsgaard_lecture_2014}. Since modes external turning points are deeper when their frequency is smaller, reconstructed time series were combined in three frequency bands: 1800-2450\,$\mu$Hz, 2450-3110\,$\mu$Hz, and 3110-3790\,$\mu$Hz following e.g. \citet{basu_thinning_2012}. VIRGO/SPM results between 1999 and 2012 are displayed in top panel of Figure~\ref{fig:figure} together with dates of cycles extrema derived from the solar radio flux\footnote{\url{ftp://ftp.seismo.nrcan.gc.ca/spaceweather/solar_flux/daily_flux_values/}.} index $\overline{F_{10.7}}$ \citep{tapping_107_2013} which is the magnetic activity index with the best correlation with seismic frequency shifts \citep{jain_solar_2009}. The high energy peaks in the combined time series are the signature of stochastic excitation of several modes at the same time. Around cycle minima, the number of high energy peaks increases (peaks higher than the arbitrary quantile at 85\% of each time series). This is reinforced in bottom panel of Figure~\ref{fig:figure} which shows
histograms of the number of high energy peaks over time in bins of $\sim1$\,year, for GOLF and VIRGO/SPM data. Additional Kolmogorov-Smirnov tests on data, corresponding to cycles maxima and minima of the combined time series, reject the null hypothesis for fully randomly excited modes during cycle minima. \citet{foglizzo_are_1998} investigated the possibility of flares and Coronal Mass Ejections (CMEs) to be a second source of excitation: flares have a suitable characteristic time but a too small strength whereas CMEs are strong enough but happen in a too large characteristic time. Furthermore, these energetic events happen more during cycle maxima which is not consistent with our observations.

\section{Conclusions}

Comparing the time series with dates of cycle extrema, we noticed that the number of high energy peaks is anti-correlated with the cycle strength: it confirms that excitation is magnetic cycle dependent, i.e. there could be a second source of excitation which could coherently excite set of modes during cycle minima. High energy flares and CMEs cannot be the second source of excitation because they occur more often at solar maximum. The higher the frequency, the more significant the variation: this suggests that the variations caused by the magnetic cycle are more pronounced close to the photosphere.
Next steps of the work is to extend the analysis to other solar-like stars observed by \textit{Kepler} to prepare PLATO (PLAnetary Transits and Oscillations of stars). 

\begin{acknowledgements}
GOLF and VIRGO instruments onboard SoHO (ESA/NASA) are a cooperative effort of many individuals to whom we are indebted. E.P. and R.A.G. acknowledge the support from the GOLF/SoHO and PLATO Centre National D’Études Spatiales grants. S.N.B acknowledges support from PLATO ASI-INAF agreement no. 2022-28-HH.0 "PLATO Fase D".

\end{acknowledgements}


\bibliographystyle{aa}  
\bibliography{Panetier_S01} 

\end{document}